# TIME AND FREQUENCY DOMAIN INVESTIGATION OF SELECTED MEMRISTOR BASED ANALOG CIRCUITS

## G. S. PATIL [a], S. R. GHATAGE [a], P. K. GAIKWAD [b], R. K. KAMAT [b], T. D. DONGALE [c, *]

[a] Gopal Krishna Gokhale College, Kolhapur, 416012, India
[b] Department of Electronics, Shivaji University, Kolhapur, 416004, India
[c] Computational Electronics & Nanoscience Research Laboratory,
School of Nanoscience and Biotechnology, Shivaji University, Kolhapur 416004, India

* Corresponding Author: tdd.snst@unishivaji.ac.in

**ABSTRACT**:

In this paper, we investigate few memristor-based analog circuits namely the phase shift oscillator, integrator, and differentiator which have been explored numerously using the traditional lumped components. We use LTspice-IV platform for simulation of the above-said circuits. The investigation resorts to the nonlinear dopant drift model of memristor and the window function portrayed in the literature for nonlinearity realization. The results of our investigations depict good agreement with the conventional lumped component based phase shift oscillator, integrator, and differentiator circuits. The results are evident to showcase the potential of the memristor as a promising candidate for the next generation analog circuits.

**Keywords:** Analog Circuits; Memristor; Simulation; Window Function.

## INTRODUCTION:

In contrast to the traditional lumped components i.e. R, L and C used for the seamless realization of any analog circuit, memristor the fourth fundamental circuit element realized by HP lab is emerging at the forefront for many interesting applications [1]. The pinched hysteresis loop in the current-voltage plane is the fingerprint characteristics of memristor which makes it distinct in comparison with the existing lumped components [2]. The intrinsic memory property makes it a strong candidate for novel applications such as non-volatile memories, biomedical appliances, programmable circuits, chaotic circuits, secure communication etc. [3-6]. The memristor fundamentally possesses two different resistance states viz. low resistance state and high resistance state. These resistance states are put to use for storing logic '0' and logic '1' for memory and logic applications [7-9]. The multilevel resistance state is yet another significant property of memristor which is useful for designing programmable analog circuits [10]. There are many important topologies of memristor-based circuit realizations; few selected ones pertinent to the present communication are covered in the following paragraph.

Recently, Muthuswamy & Kokate reported the memristor based chaotic circuit [5]. Shin et al. revealed novel application of memristor in programmable analog circuits. They have demonstrated the mid band differential gain amplifier using memristor [11]. Borghetti et al. depicted hybrid memristor-transistor logic circuit. The reported structure is useful for self-programming circuits and neuromorphic computing [12]. Robinett et al. reported the nonvolatile flip-flop circuit based on memristor device. Their reported device is compatible with the CMOS technology and efficient in terms of power and area [13]. Kim et al. reported high-performance memristor/CMOS system for the memory and neuromorphic application [14]. Talukdar et al. reported op-amp based phased shift oscillator. Their third-order memristor-based phase shift circuit produces sustained oscillation which is in good agreement with the theoretical results [15]. Adzmi et al. reported the SPICE model of memristor to design analog circuits [16].

Our research group is also active in this domain for quite some time which is evident from our few selected recent publications [3-4; 7-9; 17-24]. In the backdrop of the international scenario and based on above investigations, the present manuscript investigates the simulation of memristor-based transistorized phase shift oscillator, integrator, and differentiator circuits. The basis of these simulations is the reported literature in [15-16]. The simulations are carried out using LTspice-IV platform considering the nonlinear dopant drift model of the memristor. The rest of paper as follows, after a brief introduction, the second section discusses the simulation of transistorized phase shift oscillator. The third section reports the simulation of memristor-based integrator and differentiator circuits and at the end, the conclusion is reported.

**MEMRISTOR BASED PHASE SHIFT OSCILLATOR:**

The memristor, known for its implicit resistor with memory property has a little similarity with the simple lumped resistor. Basically, it's a passive device and hence has potential applications in the areas where lumped resistor can generally be used. In this section, we propose a memristor-based phase shift oscillator in which resistor is replaced by memristor in the phase shift arm. The present simulation is carried out using *LTspice-IV* platform. For the present simulation, the subcircuit is defined in terms of nonlinear dopant drift model of memristor and the window function reported by Biolek et al. is used for nonlinearity realization [25-26]. The mathematical formulations are as follows:

$$f(x) = 1 - (x - stp(-i))^{2p} \qquad \text{------- (1)}$$

Where,

$$stp(-i) = \begin{cases} 1, & \text{if } i \geq 0 \\ 0, & \text{if } i < 0 \end{cases}$$

Where '*p*' is a positive integer control parameter (p ∈ R⁺) and '*i*' is memristor current. The nonlinear dopant drift can be obtained by simply multiplying the state equation of memristor with the window function $f(x)$. The equation (2) presents the closed loop form of nonlinear dopant drift model of memristor using Biolek et al. window function.

$$\frac{dw(t)}{dt} = \left( \eta \frac{\mu_v R_{ON}}{D} i(t) \right) * (1 - (x - stp(-i))^{2p}) \qquad \text{------- (2)}$$

Fig. 1 reveals the memristor based phase shift oscillator. The circuit consists of three pair of Memristor-Capacitor (MC) arms which gives the total $180^0$ phase shift. Additional $180^0$ phase shift can be obtained by CE configured 2N2222 transistor. Components $R_3$ and $C_4$ give the stability against the negative feedback. Three memristor $M_1$, $M_2$, and $M_3$ with corresponding three capacitors $C_1$, $C_2$ and $C_3$ forms a feedback network and provide the necessary conditions for sustained oscillation (*Barkhausen stability criteria*).

Since the present circuit comprises of three MC arms hence can be treated as a third order dynamical system. The mathematical modeling of third order memristor-based phase shift oscillator is given in terms of Linear Time Invariant (LTI) representation and the corresponding characteristic equations for the third order phase shift oscillator system is as follows, [15]

$$as^3 + bs^2 + cs + d = 0 \qquad \text{------- (3)}$$

Where,
$$
\left.
\begin{aligned}
a &= M_1\,M_2\,M_3\,C^3\,(1 + K) \\
b &= 3M_1\,M_2\,C^2 + 2M_1\,M_3\,C^2 + M_2\,M_3\,C^2 \\
c &= 2M_1\,C + 2M_2\,C + 2M_3\,C \\
d &= 1
\end{aligned}
\right\} \qquad \text{------- (4)}
$$

For the sustained oscillation in the phase shift oscillator, the gain ($\alpha$) and frequency of oscillation ($f$) calculated is as follows,

$$w = 2\pi\,f = \frac{1}{C\sqrt{3M_1\,M_2 + 2M_1\,M_3 + M_2\,M_3}} \qquad \text{------- (5)}$$

$$\alpha = 8 + 6\frac{M_2}{M_3} + 6\frac{M_1}{M_3} + 4\frac{M_1}{M_2} + 2\frac{M_2}{M_1} + 2\frac{M_3}{M_2} + \frac{M_3}{M_1} \qquad \text{------- (6)}$$

From equations (5) and (6), results derived for the frequency of oscillation (*f*) and gain ($\alpha$), given the conditions $M_1 = M_2 = M_3 = M$ and gain ($\alpha$) as 29; the frequency of oscillation (*f*) is,

$$f = \frac{1}{2\pi MC\sqrt{6}} \qquad \text{------- (7)}$$

The equation (7) presents the formula for the sustained oscillation of phase shift oscillator [15]. Fig. 2 shows the output oscillation emanated by the phase shift oscillator. The result clearly indicates that the memristor can be used in the oscillator circuit instead of a resistor. Fig. 3 to 5 present the current and voltage behavior of memristor $M_1$, $M_2$, and $M_3$. The results clearly show that in all the individual memristors the current and voltage are in phase which is purely resistor like characteristics, and therefore in combination with a capacitor, memristor form a phase shift network and outputs sustained oscillations. It is also seen that the magnitude of the voltage and current decreases from $M_1$ to $M_3$, which is due to fact that each arm provides the loading effect on the previous stage.

Memristor exhibits resistor like characteristics at high frequency and hence it can be treated as a simple resistor [27]. Applying the same argument for memristor-based phase shift oscillator, we get three roots of characteristics equation (3); two being complex conjugate roots and other one being a purely real root [15]. The state space equation for memristor-based phase shift oscillator with a transistor in common emitter configuration has been derived as shown by equation (8). In this case, the voltage across $C_1$, $C_2$, and $C_3$ are considered as the state variables. The state space formulation for memristor-based phase shift oscillator is given as, [15]

$$\begin{pmatrix} \frac{dV_{C1}}{dt} \\ \frac{dV_{C2}}{dt} \\ \frac{dV_{C3}}{dt} \end{pmatrix} = \frac{-1}{m+1} \begin{pmatrix} \frac{1}{C_1}\left(\frac{1}{M_1} + \frac{1}{M_2} + \frac{1}{M_3}\right) & \frac{1}{C_1}\left(\frac{1}{M_1} + \frac{1}{M_2} - \frac{m}{M_3}\right) & \frac{1}{C_1}\left(\frac{1}{M_1} - \frac{m}{M_2} - \frac{m}{M_3}\right) \\ \frac{1}{C_2}\left(\frac{1}{M_1} + \frac{1}{M_2}\right) & \frac{1}{C_2}\left(\frac{1}{M_1} + \frac{1}{M_2}\right) & \frac{1}{C_2}\left(\frac{1}{M_1} - \frac{m}{M_2}\right) \\ \frac{1}{M_1 C_3} & \frac{1}{M_1 C_3} & \frac{1}{M_1 C_3} \end{pmatrix} \begin{pmatrix} V_{C_1} \\ V_{C_2} \\ V_{C_3} \end{pmatrix}$$

------- (8)

where 'm' is a normalized memristance and it is a ratio of $R_4$ and $M_1$ [15].

**MEMRISTOR BASED INTEGRATOR AND DIFFERENTIATOR CIRCUIT**

The integrator and differentiator circuits are quite established amongst the analog circuit designers due to their signal transformation capabilities. These circuits have tremendous applications in power electronics, communication, and digital electronics. A typical feature of the integrator and differentiator circuit is the *Resistor* placed in the feedback path or in the input path. Here we have introduced the memristor in the place of the resistor and simulated its response in the time and frequency domain. Fig. 6 and 7 presents the memristor based integrator and differentiator circuits respectively. In these circuits, input resistor and feedback resistor have been replaced by memristor. The modified differentiator circuit also consists of a small value resistor ($R_1$-10 Ω) in parallel with memristor to ensure proper circuit operation. Fig. 8(a-c) and 9(a-c) represents the time and frequency domain output waveform of memristor-based integrator and differentiator circuits. Fig. 8(a) shows the output response (triangular wave) of integrator circuits for square wave input signal. The triangular waveform is obtained due to the charge-discharge property of capacitor and MC time constant. The present circuit can also be used as an active low pass filter [16]. Equation (9) presents the mathematical expression of memristor-based integrator circuit and it is given as,

$$V_{OUT}(t) = -\frac{1}{MC}\int_0^t V_{IN}(t)dt + V_{OUT}(0)$$  ------- (9)

The output voltage ($V_{OUT}(0)$) is a constant of integration and it is an initial capacitor voltage at a time equal to zero (t=0). The memristor-based differentiator circuits produce the positive and negative spikes at the output stage. The output spikes are nothing but the differentiation of square wave input signal which is shown in the fig. 9(a). The equation (10) represents the mathematical expression of memristor-based differentiator circuit and it is given as,

$$V_{OUT}(t) = -MC\frac{dV_{IN}}{dt}$$  ------- (10)

Where 'M' represents the memristance. The differentiator circuits also consist of a small value resistor (10 Ω) in parallel with memristor for proper circuit operation. The value of $R_1$ is very small; hence can be neglected from the output derivation. The present circuit can also be used as an active high pass filter. The Fast Fourier Transform (FFT) results of memristor-based integrator and differentiator circuits are in good agreement with a conventional resistor based integrator and differentiator circuits. Our next aim is to develop the programmable analog circuit using thin film memristor or using memristor emulator.

**CONCLUSION:**

In conclusion, memristor-based transistorized phase shift oscillator produces sustained oscillations and other results obtained are too in good agreement with the conventional phase shift oscillator. Each MC arm produces $60^0$ phase shift with a total phase shift equal to $180^0$. The additional $180^0$ phase shift is achieved by common emitter configuration of the transistor. The time domain and frequency domain results of memristor-based integrator and differentiator are also seen to be in good agreement with the conventional integrator and differentiator circuits. Thus LRS, HRS and other finite Intermediate Resistance States (IRS) metrics of the memristor are very useful for tuning the circuit properties. The results are evident to showcase the potential of the memristor as a promising candidate for the next generation analog circuits.

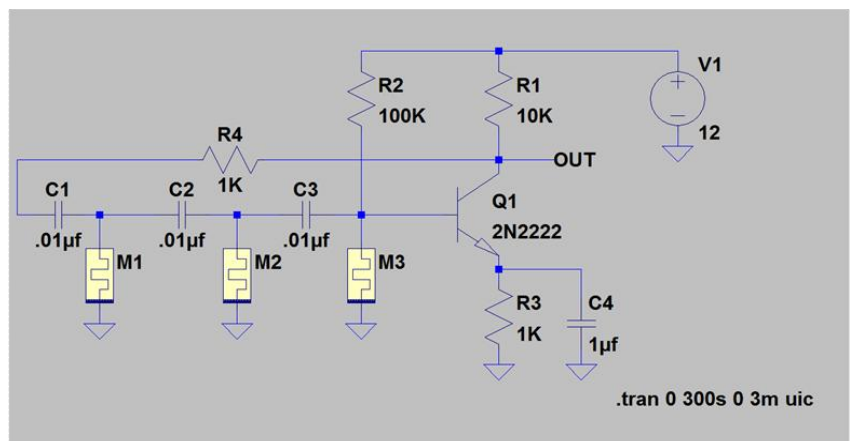

**Fig. 1:** The Memristor Based Phase Shift Oscillator. The circuit consists of three Memristor-Capacitor (MC) pairs which give total $180^0$ phase shift.

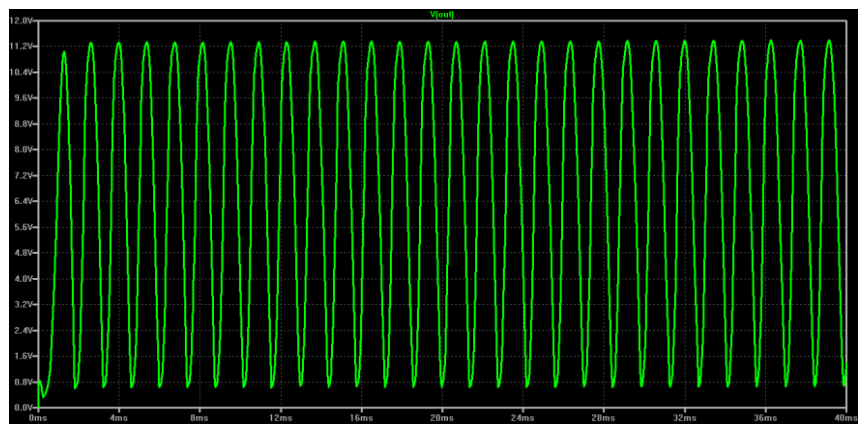

**Fig. 2:** Sustained Oscillation in the Phase Shift Oscillator.

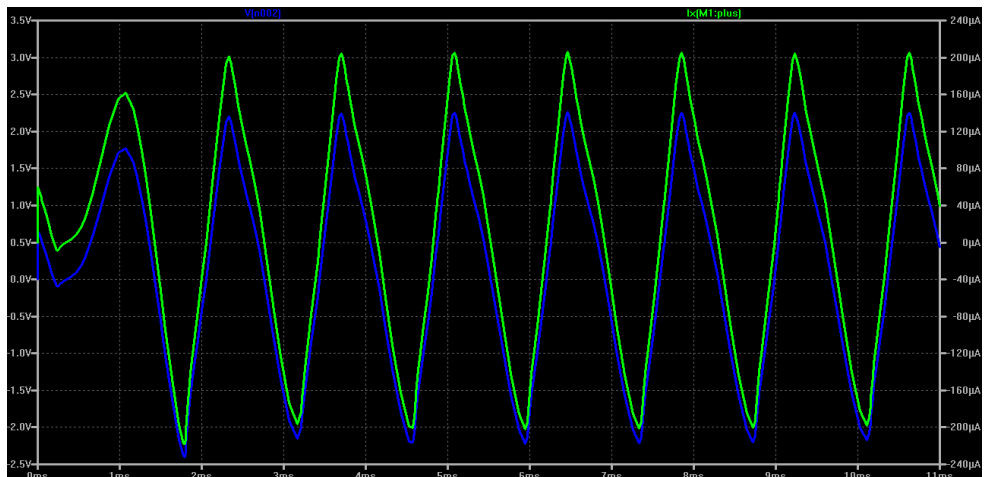

**Fig. 3:** Current and Voltage Behaviour of Memristor M₁.

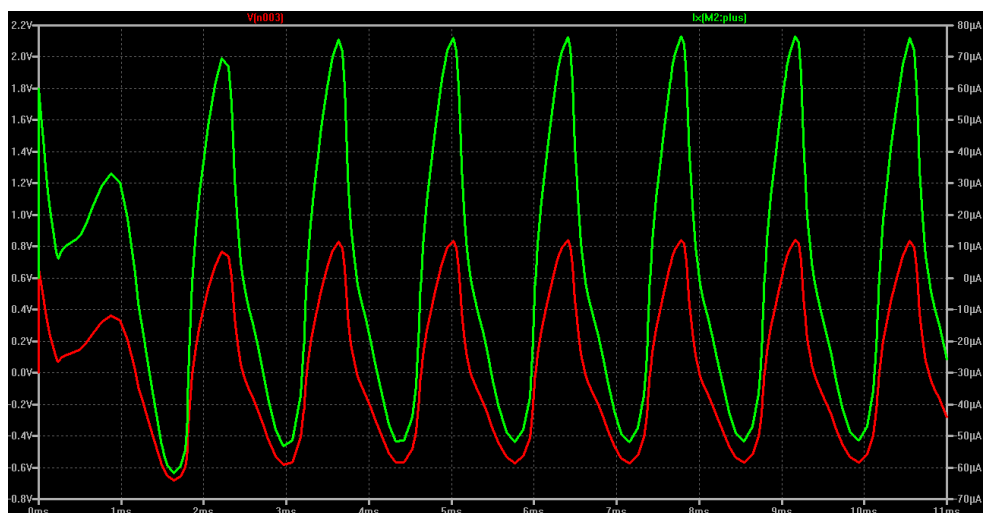

**Fig. 4:** Current and Voltage Behaviour of Memristor M₂.

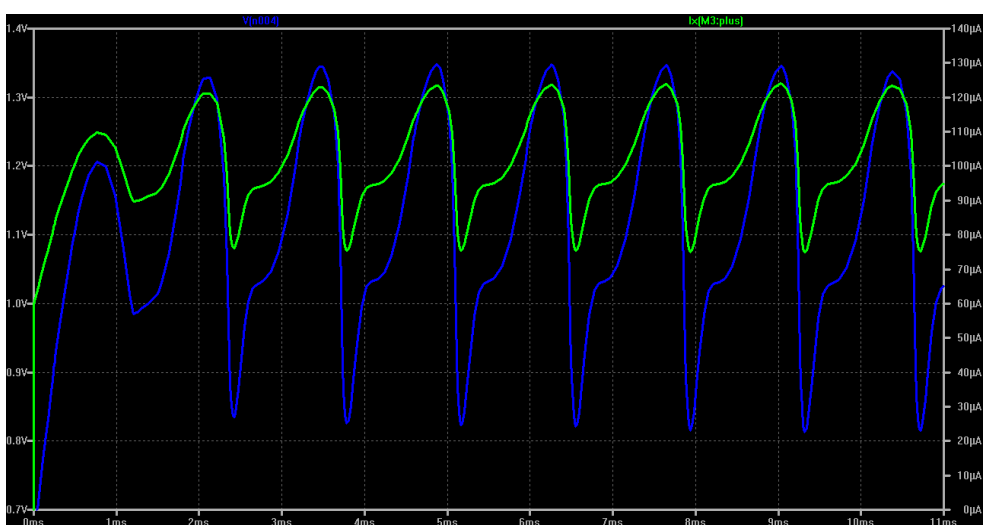

**Fig. 5:** Current and Voltage Behaviour of Memristor M₃.

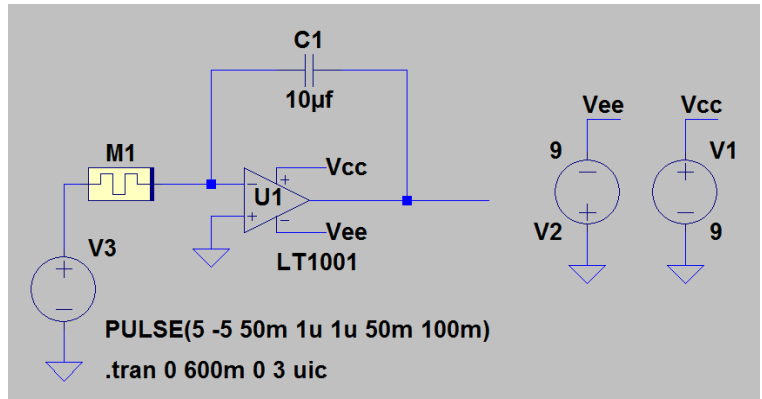

**Fig. 6:** Memristor Based Integrator Circuit

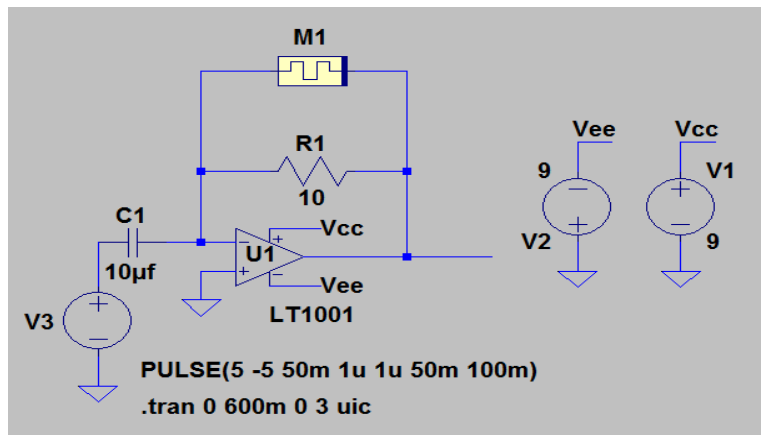

**Fig. 7:** Memristor Based Differentiator Circuit

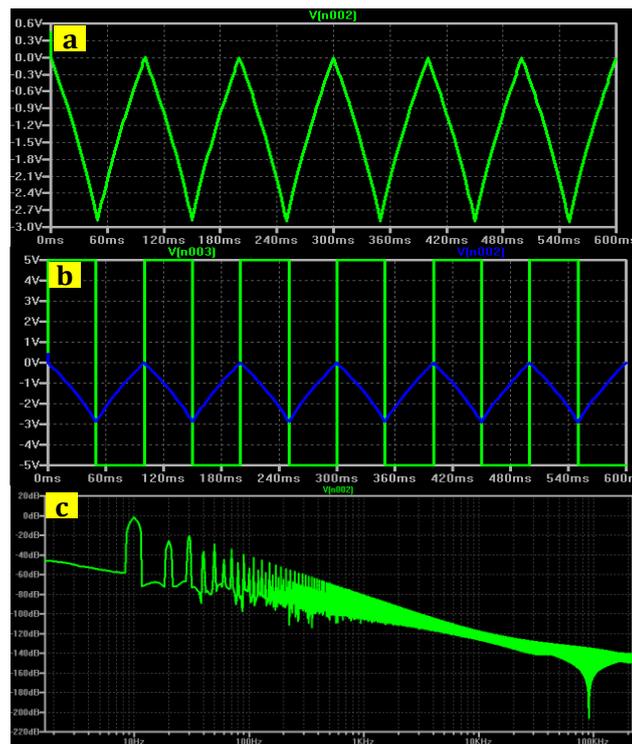

**Fig. 8:** Time and Frequency Domain Results of Memristor Based Integrator Circuits. **(a)** The triangular waveform is obtained from square wave input. **(b)** Input and output relation of square and triangular wave signals. **(c)** Fast Fourier Transform (FFT) results of memristor-based integrator circuits.

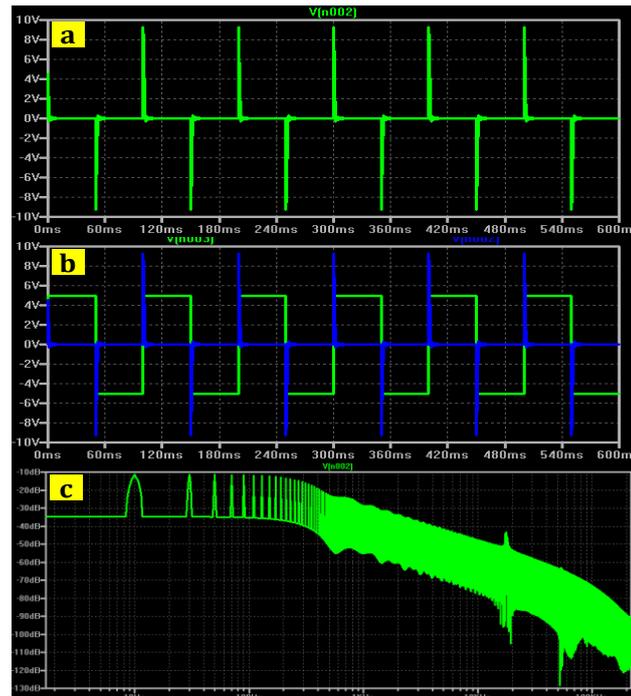

**Fig. 9:** Time and Frequency Domain Results of Memristor Based Differentiator Circuits. **(a)** The spikes waveform is obtained from square wave input. **(b)** Input and output relation of square and spikes wave signals. **(c)** Fast Fourier Transform (FFT) results of memristor-based differentiator circuits.

**REFERENCES:**

[1] L. O. Chua, *IEEE Transactions on Circuit Theory*, 18(5), pp. 507-519, 1971.

[2] M. J. Kumar, *IETE technical review*, 26(1), pp. 3-6, 2009.

[3] T. D. Dongale, S. S. Shinde, R. K. Kamat, & K. Y. Rajpure, *Journal of Alloys and Compounds*, 593, pp. 267-270, 2014.

[4] T. D. Dongale, *Health Informatics Int. J.*, 2(1), pp. 15-20, 2013.

[5] B. Muthuswamy, & P. P. Kokate, P. P., *IETE Technical Review*, 26(6), pp. 417-429, 2009.

[6] Z. Lin, & H. Wang, *IETE Technical Review*, 27(4), pp. 318-325, 2010.

[7] T. D. Dongale, K. P. Patil, S. B. Mullani, K. V. More, S. D. Delekar, P. S. Patil, P. K. Gaikwad & R. K. Kamat, *Materials Science in Semiconductor Processing*, 35, pp. 174-180, 2015.

[8] T. D. Dongale, K. P. Patil, P. K. Gaikwad, & R. K. Kamat, *Materials Science in Semiconductor Processing*, 38, pp. 228-233, 2015.

[9] S. S. Shinde, & T. D. Dongle, *Journal of Semiconductors*, 36(3), pp. 034001-1-034001-3, 2015.

[10] F. Xu-Dong, T. Yu-Hua, & W. Jun-Jie, *Chinese Physics B*, 21(9), pp. 098901-1-098901-7, 2012.

[11] S. Shin, K. Kim, & S. Kang, *IEEE Transactions on Nanotechnology*, 10(2), pp. 266-274, 2011.

[12] J. Borghetti, L. Zhiyong, S. Joseph, L. Xuemai, O. Douglas, W. Wei, D. R. Stewart, and R. S. Williams, *Proceedings of the National Academy of Sciences*, 106(6), pp. 1699-1703, 2009.

[13] W. Robinett, M. Pickett, J. Borghetti, Q. Xia, G. Snider, G. Medeiros-Ribeiro, & R. S. Williams, *Nanotechnology*, 21(23), pp. 235203-1-235203-6, 2010.

[14] K. H. Kim, S. Gaba, D. Wheeler, J. Cruz-Albrecht, T. Hussain, N. Srinivasa, & W. Lu, *Nano Letters*, 12(1), pp. 389-395, 2011.

[15] A. Talukdar, A. Radwan, & K. Salama, *Microelectronics Journal*, 43(3), pp. 169-175, 2012.

[16] A. F. Adzmi, A. Nasrudin, W. Abdullah, & S. H. Herman, In IEEE Student Conference on Research and Development (SCOReD), 2012, pp. 78-83.

[17] T. D. Dongale, K. P. Patil, S. R. Vanjare, A. R. Chavan, P. K. Gaikwad, R. K. Kamat, *Journal of Computational Science*, 11, pp. 82–90, 2015.

[18] T. D. Dongale, P. J. Patil, K. P. Patil, S. B. Mullani, K. V. More, S. D. Delekar, P. K. Gaikwad, R. K. Kamat, *Journal of Nano- and Electronic Physics*, 7(3), pp. 03012-1-03012-4, 2015

[19] T. D. Dongale, S. V. Mohite, A. A. Bagade, P. K. Gaikwad, P. S. Patil, R. K. Kamat, K. Y. Rajpure, *Electronic Materials Letters*, 11(6), pp. 944-948, 2015.

[20] T. D. Dongale, K. V. Khot, S. S. Mali, P. S. Patil, P. K. Gaikwad, R. K. Kamat, P. N. Bhosale, *Material Science in Semiconductor Processing*, 40, (2015), pp. 523–526.

[21] T. D. Dongale, P. S. Pawar, R. S. Tikke, N. B. Mullani, V. B. Patil, A. M. Teli, K. V. Khot, S. V. Mohite, A. A. Bagade, V. S. Kumbhar, K. Y. Rajpure, P. N. Bhosale, R. K. Kamat, P. S. Patil, *Journal of Nanoscience and Nanotechnology*, 17, pp. 1–8, 2017. DOI:10.1166/jnn.2017.14264

[22] T. D. Dongale, K.V. Khot, S.V. Mohite, S.S. Khandagale, S.S. Shinde, V.L. Patil, S.A. Vanalkar, A.V. Moholkar, K.Y. Rajpure, P.N. Bhosale, P.S. Patil, P.K. Gaikwad, R.K. Kamat, *Journal of Nano- and Electronic Physics*, 8(4), pp. 04030-1-04030-4, 2016.

[23] T. D. Dongale, N. D. Desai, K. V. Khot, N. B. Mullani, P. S. Pawar, R. S. Tikke, V. B. Patil, P. P. Waifalkar, P. B. Patil, R. K. Kamat, P. S. Patil, P. N. Bhosale, *Journal of Solid State Electrochemistry*, DOI: 10.1007/s10008-016-3459-1

[24] T. D. Dongale, P. J. Patil, N. K. Desai, P. P. Chougule, S. M. Kumbhar, P. P. Waifalkar, P. B. Patil, R. S. Vhatkar, M. V. Takale, P. K. Gaikwad, R. K. Kamat, *Nano Convergence*, 3(1), pp. 1-7, 2016.

[25] Z. Biolek, D. Biolek, & V. Biolkova, *Radioengineering*, 18(2), pp. 210-214, 2009.

[26] S. Kvatinsky, E. G. Friedman, A. Kolodny, and U. C. Weiser, *IEEE Transactions on Circuits and Systems—I: Regular Papers*, 60(1), pp. 211-221, 2013.

[27] Y. N. Joglekar, & S. J. Wolf, *European Journal of Physics*, 30(4), pp. 661-675, 2009.